# Sparse Nonnegative Matrix Factorization for Multiple Local Community Detection

Dany Kamuhanda, Meng Wang, Kun He*, *Senior Member, IEEE*

***Abstract*— Local community detection consists of finding a group of nodes closely related to the seeds, a small set of nodes of interest. Such group of nodes are densely connected or have a high probability of being connected internally than their connections to other clusters in the network. Existing local community detection methods focus on finding either one local community that all seeds are most likely to be in or finding a single community for each of the seeds. However, a seed member usually belongs to multiple local overlapping communities. In this work, we present a novel method of detecting multiple local communities to which a single seed member belongs. The proposed method consists of three key steps: (1) local sampling with Personalized PageRank (PPR); (2) using the sparseness generated by a sparse nonnegative matrix factorization (SNMF) to estimate the number of communities in the sampled subgraph; (3) using SNMF soft community membership vectors to assign nodes to communities. The proposed method shows favorable accuracy performance and a good conductance when compared to state-of-the-art community detection methods by experiments using a combination of artificial and real-world networks.**

## I. INTRODUCTION

Many complex data such as user interaction in social networks, product purchases, scientific collaboration, interaction among proteins of an organism are represented by a graph (network) consisting of nodes connected by edges indicating the interactions [1]. There are three tasks that dominate the area of network analysis: (1) node classification that consists of predicting the label of a target node based on other labeled nodes [2] [3]; (2) link prediction that consists of predicting an edge between two unconnected nodes [4] [5]; (3) community detection that finds a set of closely related nodes within a network [6] [7]. We are interested in community detection as this task can also be used to address the other two tasks. For node classification, a node can be assigned a label carried by its community members; and for link prediction, nodes of the same community are more likely to connect with each other as compared to nodes from different communities.

A community is a group of nodes that are densely connected (cluster) or have a high probability of being connected internally than their connections to the nodes in other clusters of the network [8]. Community detection algorithms include: (1) global algorithms to detect all communities on the entire network [9] [10] [11] [12] [13] [14]; (2) local algorithms that find a single local community to which a seed set belongs [15] [16] [17] [18] [23] [24]; (3) local algorithms that find multiple local communities around a seed set [19] [20] [21].

Community detection addresses a number of real-world problems. In social networks, one may be interested in finding all groups of family members or friends in a network for a specific reason such as epidemic control. In communication networks, a suspect group linked to known criminals is more likely to be identified. Community detection can also be used for feature selection to speed up a machine learning algorithm [22], or find customer shopping patterns in recommendation systems.

As real-world networks (such as Facebook) are very large with millions or billions of nodes, detecting all communities globally becomes a computationally expensive task. Moreover, most of the time we are only interested in a subset of nodes of a local region. In such cases, the local community detection provides a solution [16] [23] [24]. For instance, a chef may know a few ingredients used for a particular recipe but these seeding ingredients are not available on the market. Using a flavor network, local community detection methods can identify similar ingredients that could replace these ingredients without having to detect all communities of the flavor network [20] [25]. Likewise, recommender systems can use the similarity among consumers with similar shopping views or purchase same products to recommend relevant products to a consumer [26]. Single local community detection assumes that all seeds belong to the same community and the task is to find missing members of the community. Bian et al. [20] assume that multiple seeds may belong to different communities and find a single community for each seed, which becomes a single local community detection task for each seed.

We address a more challenging problem that is rarely addressed in the literature. Given a single seed, the task is to find all possible communities that the seed belongs to. Compared with the multiple local community detection, the problem of finding one single community a seed is most

Manuscript received: 2020-01-19

This work is supported by National Natural Science Foundation (61772219) and the Fundamental Research Funds for the Central Universities (2019kfyXKJC021).

D. Kamuhanda is with the School of Computer Science and Technology, Huazhong University of Science and Technology, Wuhan, China, 430074 (e-mail: kamuhanda@hust.edu.cn).

M. Wang is with the School of Computer Science and Technology, Huazhong University of Science and Technology, Wuhan, China, 430074 (e-mail: mengwang233@hust.edu.cn).

K. He is with the School of Computer Science and Technology, Huazhong University of Science and Technology, Wuhan, China, 430074. Corresponding author (e-mail: brooklet60@hust.edu.cn).

likely to be in is easier as this community is denser or stronger than other local communities this seed is in. The multiple local community detection problem is specifically difficult because we cannot use more than one seed to improve the accuracy as in other existing local community detection tasks. For this task, even when we know some nodes that belong to the same community as the seed, they are not helpful for improving the accuracy since they do not belong to all communities of the seed. In addition, as nodes belong to multiple communities, the overlapping portion becomes very dense and may sometimes look like a single community which is difficult to be split into different communities. The problem of multiple community detection for a single seed was first introduced in [21] but the authors focused more on single local community detection. In our previous conference publication [19], we focused on the multiple local community detection for a single seed based on nonnegative matrix factorization (NMF). This work is an extended version with significant improvement on [19].

In this work, we propose a Sparseness-based Multiple Local Community detection method (S-MLC) for finding multiple local communities of a single seed. There are three key steps in S-MLC: local sampling to find relevant nodes for the seed; estimating the number of communities in the sampled subgraph; and detecting the local communities. We use SNMF to learn the structural information of a network as SNMF can find better representations than NMF [27][28]. We then use soft community membership vectors generated by SNMF to assign nodes to their corresponding communities. Other network embedding methods such as those based on graph neural networks (GNNs) [29] [30], Node2Vec [31], DeepWalk [32] require the input node features which are not always available or generate embeddings which are difficult to interpret. The nonnegativity feature of SNMF allows each node's embedding vector to be interpreted as probabilities of belonging to different communities and makes it more suitable for this task.

S-MLC uses similar framework of our conference version of MLC [19] in terms of local sampling and community estimation. MLC uses a BFS for local sampling while S-MLC uses Personalized PageRank (PPR), which is often used for local sampling in existing single local community detection methods [24]. PPR is computationally expensive but efficient approximations [15] [17] can be used. To estimate the number of communities, MLC iterates NMF decompositions until the normalized $H$ (values of each column sum up to 1) contains a row without a centroid node. S-MLC estimates communities by iterating SNMF on the sampled subgraph where the number of components that yield the maximum sparseness is used as the number of communities. This approach is based on the sparse coding which aims to find few elements that can effectively represent the entire population. Regarding the final phase of community detection, both S-MLC and MLC use a threshold on community membership vectors generated by NMF/SNMF algorithm to assign nodes to communities.

Our contributions are summarized as follows. First, we propose a method of estimating the number of communities in social networks, which can be used in other community detection algorithms that require the number of communities as prerequisites. Second, we investigate and conduct extensive experiments on local sampling techniques to determine which techniques and parameters are suitable for sampling a subgraph containing almost all members of multiple local communities of a given seed. Specifically, we demonstrate the sampling behaviour of PPR and HK which is useful in making a good choice between the two methods depending on the application scenario. Third, we address a challenging problem of finding multiple local communities for a single seed, which is rarely addressed in the literature, and propose a novel method, called S-MLC for solving this problem. The proposed approach outperforms the state-of-the-art baselines via extensive experiments. S-MLC outperforms MLC [19] and M-LOSP [21] as evaluated on artificial and real-world networks. As there are few algorithms for multiple local community detection, we further run DEMON on the sampled subgraph to find approximate local communities, denoted as L-DEMON, and both S-MLC and MLC clearly outperform L-DEMON on most networks.

The rest of the paper is organized as follows. Section II discusses the related work. Section III introduces the addressed problem, measures of evaluating the solution quality, local sampling with focus on PPR and how NMF is used for community detection. Section IV discusses the three steps of the proposed method for detecting communities of a single seed. Section V presents experimental results, followed by a conclusion.

## II. RELATED WORK

### A. *Local Community Detection*

There are two types of problem formulation for the local community detection: (1) assuming all seeds belong to one community; (2) assuming the seeds belong to multiple communities simultaneously.

Clauset [33] and Chen et al. [34] iteratively expand a community $C$ which initially consists of a seed set by adding more members one-by-one from its boundary. A node is removed from the boundary and added to $C$ if it improves the local modularity [33] or the internal relation [34], then its neighbors join the boundary. The advantage of this approach is that we can query and get current community members at any time even when the algorithm is still processing. Personalized PageRank (PPR) [15] [35] and Heat Kernel (HK) [17] based methods also assume that all seeds belong to the same community, then consider the seed set as an initial community and grow the community by sorting their probability vector $p$ in the decreasing order to obtain $q$ followed by finding a set of nodes with a minimum conductance. The advantage of these approaches is that, the more initial seeds we have, the higher the accuracy of the output. Spectral-based methods such as LOSP [21] [24] find a local spectral basis supported by the seeds through a few steps of random walks. Then a sparse vector $y$ is computed and contains probabilities indicating the extent to which each node is likely to belong to the same community as the seed set. He et al. [24] systematically build a family of local spectral subspace-based methods through various diffusions from the seeds to form the local spectral subspace.

He et al. [21] put a section to extend their single local community detection algorithm called LOSP to find all communities a single seed belongs to, which is the first work to address the multiple local community detection problem. They temporarily remove the seed from its ego network to get connected components, then for each connected component they add the seed back to build an initial seed set and use LOSP as a subroutine to find local community for each initial seed set. This approach allows to detect multiple local communities and we denote it as M-LOSP. Our previous work of MLC [19] uses a few steps of breadth-first search for local sampling, and then uses a nonnegative matrix factorization (NMF) to learn the network structure encoded in the adjacency matrix so that nodes can be assigned to communities based on a threshold applied to community membership vectors generated by NMF. Hollocou et al. [36] use local scoring metrics (PPR [15], Heat Kernel [17] and LEMON [23] score) to define an embedding of the graph around the seed set and pick new seeds based on the embedding to uncover multiple communities, but the newly found communities are not for the original seeds but for the newly determined seeds.

### *B. Network Embedding*

Recent community detection methods consist of encoding a network so that nodes belonging to the same community yield similar representation compared to nodes belonging to different communities. Such a representation is obtained by learning the network features using SVD [37], NMF [19], Autoencoder [38], DeepWalk [32], Node2Vec [31] or graph neural network (GNN) [3] [30].

Abdollahpouri et al. [39] encodes the input network using a set of locus-based adjacency representations where each representation consists of $n$ nodes each linked with a random neighbor. Connected components of each representation become communities, generating a solution by the representation. Solutions generated by each representation are compared by optimizing both the Kernel K-means (KKM) and Ratio Cut (RC) and eligible solutions are retained. Among these retained solutions, the best is the one with a good normalized mutual information (NMI). This method has an advantage of not requiring the number of communities. However, it is applicable to global community detection and not for local community detection.

Mahmood et al. [40] generate a new representation of a network by computing geodesic distances of nodes, then use a sparse linear coding to decompose the obtained representation into a matrix of coefficients which is clustered using spectral clustering to obtain communities.

NMF has an advantage of obtaining communities without requiring an additional clustering method and its nonnegativity feature makes it easy to detect both overlapping and non-overlapping communities [19]. Recent NMF-related works detect global communities by adding a regularization term that incorporates local and global network information for improving the accuracy [41] [42]. The representations obtained using other mentioned methods are mainly used for non-overlapping community detection or some other tasks like link-prediction or node classification. Specifically, GNNs are suitable for networks with node labels in semi-supervised tasks where available labels can be used in predicting unknown labels. Non-overlapping communities can be detected using a clustering algorithm such as *k*-means with an input *k* as the number of communities and a representation of the network. The quality of detected communities not only depend on the type of network representation but also the choice of the community number.

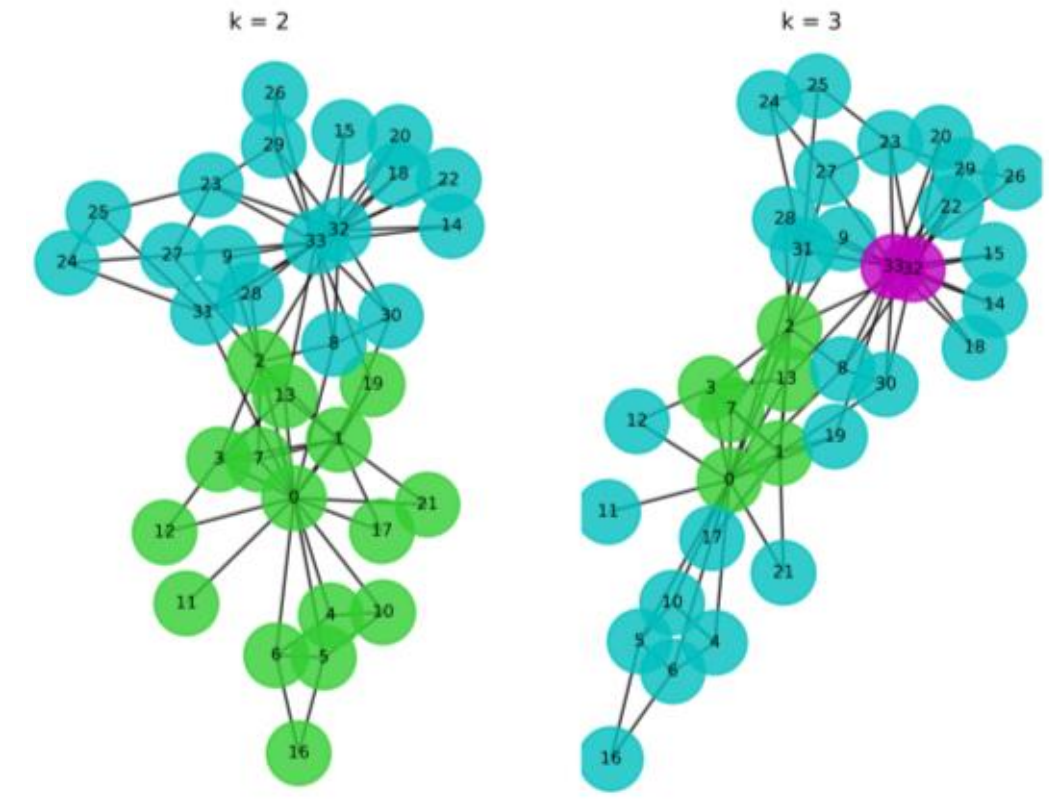

Fig. 1. Clusters found by *k*-means for different values of *k* on the *Zachary Karate club* network. The choice of the community number greatly affects the quality of community detection.

### *C. Estimating the Number of Communities*

Eigengap is a heuristic to estimate the number of clusters where the estimated number of clusters *k* is related to the largest gap between consecutive eigenvalues $\lambda_k$ and $\lambda_{k+1}$ of a Laplacian matrix and small gaps between $\lambda_1, \lambda_2 \dots \lambda_k$ [43][37]. In some cases, such as in networks with overlapping communities, there is no clear gap and eigengap heuristic does not work [43].

Modularity maximization model [44] is the most popular method for estimating the number of communities. It is based on a modularity matrix which encodes the eigenvectors of a network and is calculated as:

$$B_{ij} = A_{ij} - \frac{\mathrm{d}(v_i)\mathrm{d}(v_j)}{2m},$$

where *A* represents the adjacency matrix, $\mathrm{d}(v_i)$ the degree of a node $v$ of the $i^{\text{th}}$ index in *A*, and *m* the total number of edges in the network. For a network of two disjoint communities with $h_i \in \{1, -1\}$, the community membership indicator for a node $v_i$, namely the modularity is computed as [44]:

$$Q = \frac{1}{4m} h^T B h,$$

where $h$ is the column vector whose elements are $h_i$. The decomposition of a network into more than two communities can be done hierarchically by dividing the network into two communities, then each of them divided into two sub-communities and so on up to non-increasing modularity. As $B$ is symmetric, there exists a decomposition such that $B \cong U \Lambda U^T$ where $U$ is a matrix of eigenvectors and $\Lambda$ is a matrix of eigenvalues on the diagonal. By extracting some eigenvectors from $U$, we have $H$ that corresponds to the top $k$ eigenvalues in $\Lambda$, sorted in a non-increasing order. The generalized modularity for more than two communities can be computed as [38]: $Q = \sum_{ii} H^T BH$. The final number of communities in a network is the one that maximizes modularity $Q$ from various iterations of $k$.

## III. PRELIMINARIES

### *A. Problem Formulation*

Given a network modelled as a graph $G = (V, E)$ and a seed $s \in V$ where $G$ is an undirected, unweighted graph with $n$ nodes $V = \{v_1 \cdots v_n\}$ and $m$ edges $E = \{e_1 \cdots e_m\} \subseteq V \times V$, an adjacency matrix $A \in \mathbb{N}^{\,n\,x\,n}$ can be constructed to represent the network such that entry $a_{ij} = 1$ if a node $v_i$ is connected to $v_j$ and otherwise $a_{ij} = 0$. Let $C^{(s)}$ be the set of $k$ $(k \geq 1)$ ground-truth communities containing $s$. The problem of concern is to detect communities $C'^{(s)}$ for seed $s$ such that

$$\forall\, C_i^{(s)} \in C^{(s)} \Longrightarrow \exists\, C_j'^{(s)} \in C'^{(s)} \mid C_i^{(s)} \equiv C_j'^{(s)},$$

where $1 \leq i \leq k$ and $\left|C'^{(s)}\right| = k' \cong k$. Here $|\cdot|$ indicates the number of communities.

To reduce the computational complexity, we usually search the communities in a sampled subgraph $G_s = (V_s, E_s) \subseteq G$. In the process of addressing the main problem, we consider several sub-problems: local sampling and estimating the number of communities in $G_s$. The number of communities is important as it affects the quality of follow-up community detection and clustering algorithms. For example, Fig. 1 shows how $k$-means clustering finds accurate communities in the *Zachary karate club* network [45] for $k = 2$ (its ground-truth community number) than the clustering for $k = 3$.

### *B. Evaluation Measures*

A number of measures have been selected to evaluate the quality of solutions to the addressed problem.

The *conductance* of a detected community $C_j'^{(s)}$ is a fraction of its edges that points outside the community[46] :

$$\emptyset\left(C_j'^{(s)}\right) = \frac{\sum_{\left(i \in C_j'^{(s)},\ j \notin C_j'^{(s)}\right)} A_{ij}}{min\left(Vol\left(C_j'^{(s)}\right),\ \left(Vol(V) - Vol\left(C_j'^{(s)}\right)\right)\right)}, \qquad (1)$$

where $Vol(\cdot)$ denotes the total degree of a set of nodes. The lower the value, the denser the inside connections with sparser outside connections for the community.

The *recall* of a ground-truth community $C_i^{(s)}$ indicates how well it has been detected:

$$Rec\left(C_i^{(s)}\right) = max_{j=1\cdots k'} \frac{\left|C_i^{(s)} \cap C_j'^{(s)}\right|}{\left|C_i^{(s)}\right|}. \qquad (2)$$

The *precision* of a detected community $C_j'^{(s)}$ indicates its relevance compared to the ground-truth communities:

$$Prec\left(C_j'^{(s)}\right) = max_{\,i=1\cdots k} \frac{\left|C_i^{(s)} \cap C_j'^{(s)}\right|}{\left|C_j'^{(s)}\right|}. \qquad (3)$$

A combined measure $F_\sigma$ $(\sigma = 1,2)$ can be computed where $F_1$ balances the *precision* and *recall*, and $F_2$ focusses more on the *recall* [47]:

$$F_\sigma = (1 + \sigma^2) \times \frac{Prec \times Rec}{(\sigma^2 \times Prec) + Rec}. \qquad (4)$$

### *C. Graph Diffusion*

Graph diffusion consists of spreading a node mass step by step throughout the graph and this approach is very popular in local community detection [20] [21] [15] [17]. For a seed $s$, an initial vector $p^{(0)} \in \mathbb{R}^n$ consisting of ones for the seed and zeros for the remaining nodes indicating the distribution of a random walker at the initial stage. Then the random walker spreads information across the graph starting from the neighbors of $s$. At the first iteration, neighbors of $s$ have $1/d(s)$ probability of being visited, and zero for the remaining nodes, where $d(s)$ is the degree of node $s$. For the entire network, this information can be summarized into a transition matrix $T$ where $T_{ij} = A_{ij}/d(i)$ or $T = D^{-1}A$ where $D$ is the diagonal degree matrix. The random walk step $k$ consists of computing new probability vectors [15]:

$$p^{(1)} = Tp^{(0)}, \cdots, p^{(l)} = Tp^{(l-1)}.$$

PPR [15], initially proposed by Brin and Page [48] for ranking webpages, uses this idea to rank nodes based on the seed with $\alpha$ probability of following any edge and $(1 - \alpha)$ probability to restart:

$$p^{(l)} = \alpha * Tp^{(l-1)} + (1 - \alpha) * p^{(0)}, \qquad (5)$$

where $\alpha \in (0,\ 1)$. Andersen *et al.* [15] proposed a fast approximation of $p$ that finds a probability vector $p'$ using push operations. It uses two vectors $p'$ and $r$ (nonnegative) and each push operation copies some probability from $r$ to $p'$ while the remaining probability gets spread in $r$. The adapted pseudocode is provided in Algorithm 1 while the original pseudocode can be found in [15].

**Algorithm 1: ApproximatePPR**

```
INPUT: G, s, α, ε
Initialize:
 r[s] = 1 #start with the highest probability on seed s
 Q = [s] #queue to store any node x if r[x] ≥ εd(x)
 p' = { } #start with an empty sample
while (Q is not empty) do:
    u = Q.dequeue #first in first out
    if u not in p' do: p'[u] = 0 end if
    p'[u]+= (1 − α) * r[u] #copy 1−α probability to p'
    remaining = α * r[u] #remaining probability
    r[u] = remaining/2 #keep a half of it in r and
    spread the remaining to the neighbors
    for v in neighbors of u do:
        if v not in r do: r[v] = 0 end if
        if r[v] < εd(v) and r[v] + remaining/(2*d(u)) ≥ εd(v)
            do: Q.append(v) end if
        r[v] = r[v] + remaining/(2*d(u))
    if r[u] ≥ εd(u) do: Q.append(u) end if
RETURN p'
```

Another graph diffusion is Heat Kernel (HK) which is a function of temperature $t$ and the initial heat distribution $h^{(0)}$

[49]: $h = H^{(t)}h^{(0)}$, where $H^{(t)}$ is the heat operator: $H^{(t)} = e^{-tT}$. As the exponent of any matrix $e^{A} = \sum_{k=0}^{\infty}\frac{A^k}{k!}$, the heat kernel diffusion becomes [49]:

$$h = e^{-t}\left(\sum_{k=0}^{\infty}\frac{t^k}{k!}T^k\right)h^{(0)} . \quad (6)$$

And Kloster et al. [17] proposed a fast approximation of HK (HK-Relax), which at each step $j$ copies a probability from a residual $r$ for a node $v$ to $p'$ if $r(v,j) \geq \frac{e^t\varepsilon d(v)}{2N\psi_j(t)}$ [17]. Initially, for a seed $s$ of the seed set $S$ the residual $r(s,0) = 1/|S|$.

### D. NMF

NMF is a matrix factorization technique which reduces an input nonnegative matrix $A \in \mathbb{R}^{m\times n}$ to two nonnegative matrices $W \in \mathbb{R}^{m\times k}$ and $H \in \mathbb{R}^{k\times n}$ such that $A \cong WH$ [50]. The number of components $k$ should be specified. NMF consists of solving one of the following optimization problems [51]:

$$\min_{W,H\geq 0} J_F(W,H) = \|A - WH\|_F^2 , \quad (7)$$

$$\min_{W,H\geq 0} J_{KL}(W,H) = \sum_{ij}\left(A_{ij}\log\frac{A_{ij}}{[WH]_{ij}} - A_{ij} + [WH]_{ij}\right), \quad (8)$$

where $J_F$ and $J_{KL}$ denote the Frobenius-norm and Kullback-Leibler divergence cost functions. The optimization is done using the multiplicative update rules in Eq. (9) and Eq. (10) or Eq. (11) and Eq. (12) respectively [51]:

$$H_{kj} \leftarrow H_{kj}\frac{(W^TA)_{kj}}{(W^TWH)_{kj}} , \quad (9)$$

$$W_{ik} \leftarrow W_{ik}\frac{(AH^T)_{ik}}{(WHH^T)_{ik}} , \quad (10)$$

$$H_{kj} \leftarrow H_{kj}\frac{\sum_i W_{ik}A_{ij}/(WH)_{ij}}{\sum_i W_{ik}} , \quad (11)$$

$$W_{ik} \leftarrow W_{ik}\frac{\sum_j H_{kj}A_{ij}/(WH)_{ij}}{\sum_j H_{kj}} . \quad (12)$$

The detection of communities with NMF is done by normalizing $H$ using Eq. (13) to generate the community membership probabilities of the nodes:

$$H_{ij} = \frac{h_{ij}}{\sum_{x=1}^{k} h_{xj}} . \quad (13)$$

Hard clustering of nodes can be done based on the highest membership value [12], and overlapping communities can be detected by assigning nodes to multiple communities if more than one of its community membership values exceed a particular threshold [14] [19]. We use the sparseness generated by a sparse NMF to estimate the number of communities.

## IV. THE S-MLC APPROACH

The proposed S-MLC approach includes three steps, as illustrated in Fig. 2.

### A. Local Sampling

The objective of local sampling is to maximize the recall but also balance it with the precision. A high recall indicates that most of the nodes we want have been sampled, and a high precision indicates that we have few irrelevant nodes that need to be discarded during the community detection phase. We use PPR or HK approximate algorithms [15][17]. For PPR, we fix $\alpha = 0.99$ and vary $\varepsilon \in \{10^{-3}, 10^{-4}\}$. The smaller $\varepsilon$ is, the larger the sample as the algorithm stops when the probability of every node $v$ in $V$ is less than $\varepsilon d(v)$. More details can be found in [15]. For HK, we vary $\varepsilon \in \{10^{-2}, 10^{-3}\}$ and $t \in \{80, 40\}$ respectively then compute $N = 2t * log(1/\varepsilon)$ and $psis$ ($\psi$) automatically as discussed in [17].

We first construct a subgraph $subG_s$ consisting of nodes that are associated with positive probabilities in $p'$. Then, we find its biconnected components using a popular algorithm proposed by Hopcroft et al [52] and sort them based on their size. A top (the largest) biconnected component containing $s$ ($G_s$) is used as a sample. This means that every node in $G_s$ can be reached from any node in $G_s$ and the removal of any node does not disconnect $G_s$. The main reason of using such a biconnected component is based on the core-periphery structure [53] [46] of network communities in which overlapping nodes tend to form a dense core with whiskers (less connected) around. The biconnected component returns almost all members of the communities of $s$.

**Algorithm 2. Local Sampling**

1. **INPUT**: Graph $G = (V, E)$ and seed $s \in V$
2. **Set** PPR approximation parameters: $params$
3. $p' =$ **ApproximatePPR**$(G, s, params)$
4. $V_s = [v$ for $v$ in $V$ if $p'[v] > 0]$
5. $subG_s =$ **subgraph**$(G, V_s)$
6. $G_s =$ biggest biconnected component in $subG_s$ containing $s$
7. **RETURN** $G_s$

### B. Estimating the Number of Communities

Local sampling usually includes nodes that are not part of the communities of the seed $s$. Even when local sampling returns nodes that match those in the ground-truth communities of $s$, we still need to assign each node to its community as $s$ may belong to multiple communities. We tackle this problem by first estimating the number of communities in the sampled subgraph based on a sparse coding that aims to find a few elements that can represent the entire population [54].

Given $l$ input vectors $\{h\}_{i=1}^{l} \in \mathbb{R}^n$, sparse coding aims to find sparse codes $\{h_i\}_{i=1}^{l} \in \mathbb{R}^k$ and a dictionary $W \in \mathbb{R}^{n\times k}$ (basis vectors) such that each $h_i$ can be found using a linear combination of $w^j$: $h_i = \sum_{j=1}^{k} h_i^j w^j = Wh_i$ [54]. Most of the coefficients $h_i^j$ are zeros or close to zero with few values (representative) far from zero, resulting in a sparse vector $h_i$. By using NMF, the sparsity can be enforced by adding penalties on either $W$, $H$ or both as follows:

$$J(W,H) = J_z(\mathrm{A}, \mathrm{WH}) + \omega\Phi(W) + \beta\psi(H) , \quad (14)$$

where $J_z$ is a cost function in Eq. (7) or Eq. (8); $\omega, \beta$ are sparsity parameters; $\Phi, \psi$ are sparsity functions. $L_0$ norm would be the most appropriate but it is difficult to optimize due to the non-differentiability. $L_1$ norm is the most popular sparsity function and has been used in many applications with good results [27] [28].

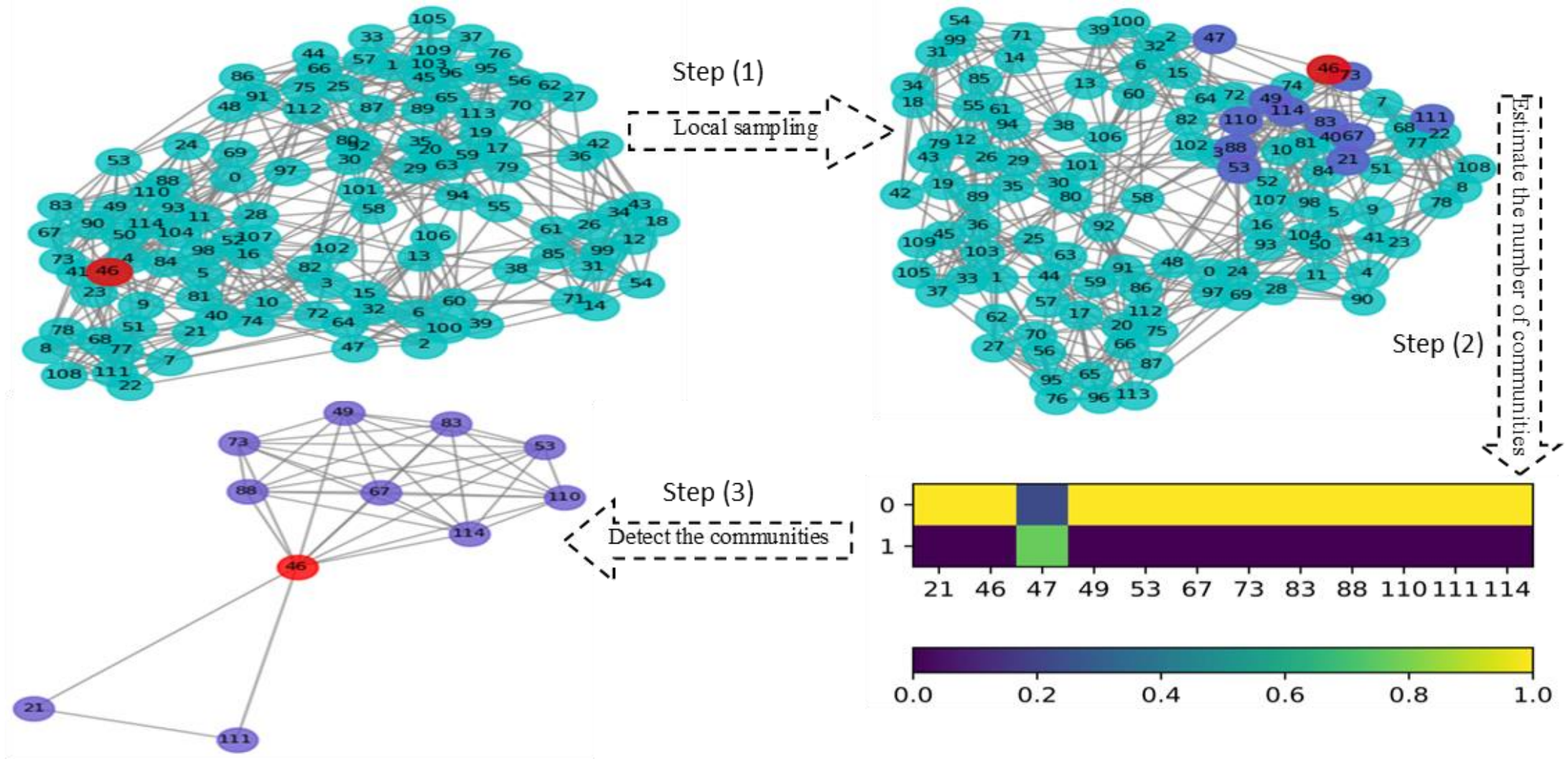

Fig. 2. Local community detection steps. We start with a seed *s* (node 46), sampling nodes around *s*, then estimate the number of communities in the sampled subgraph and detect each of them based on the learned structure. The color bar indicates the community membership probabilities in *H*. Notice how the node 47 is discarded as it forms its own community compared to other sampled nodes.

The sparsity can be imposed on rows of $H$ and/or columns of $W$. For community detection, the sparsity on rows of $H$ would activate few nodes in each community (the nodes belonging to the community) which is only useful when we know the number of communities.

In case we do not know the number of communities, the sparsity on columns of $H$ would activate a few communities (components) for each node. In other words, each node belongs to a few communities as opposed to the sparsity on rows which assumes that each community consists of a few nodes.

We use the SNMF on columns proposed by Kim *et al.* [28] to iteratively decompose the adjacency matrix of $G_s$ and estimate the number of communities based on the sparseness of $H$:

$$\min_{W,H\geq 0}\left\{\|A-WH\|_F^2+\ \beta\sum_{j=1}^{n}\left\|h_j\right\|_1^2\right\}. \tag{15}$$

The optimization of Eq. (15) is done by alternating Eq. (16) and Eq. (17) [28]:

$$\min_{W\geq 0}\|H^TW^T-A^T\|_F^2\ , \tag{16}$$

$$min_{H\geq 0}\left\|\begin{pmatrix}W\\ \sqrt{\beta}\mathbf{1}\end{pmatrix}-\begin{pmatrix}W\\ \mathbf{0}\end{pmatrix}\right\|_F^2, \tag{17}$$

where $\mathbf{1}$ is an all one row vector of dimension $1\times k$ and $\mathbf{0}$ is an all-zero row vector of dimension $1\times n$.

SNMF generates partitions that overlap in which every node is associated with a probability value of belonging to each community, and this is useful in detecting overlapping communities of the seed. In addition, the sparsity property, when applied to columns of coefficients, activates a few components for each node and demonstrates the total number of communities in the network (the highest number of components that have been activated).

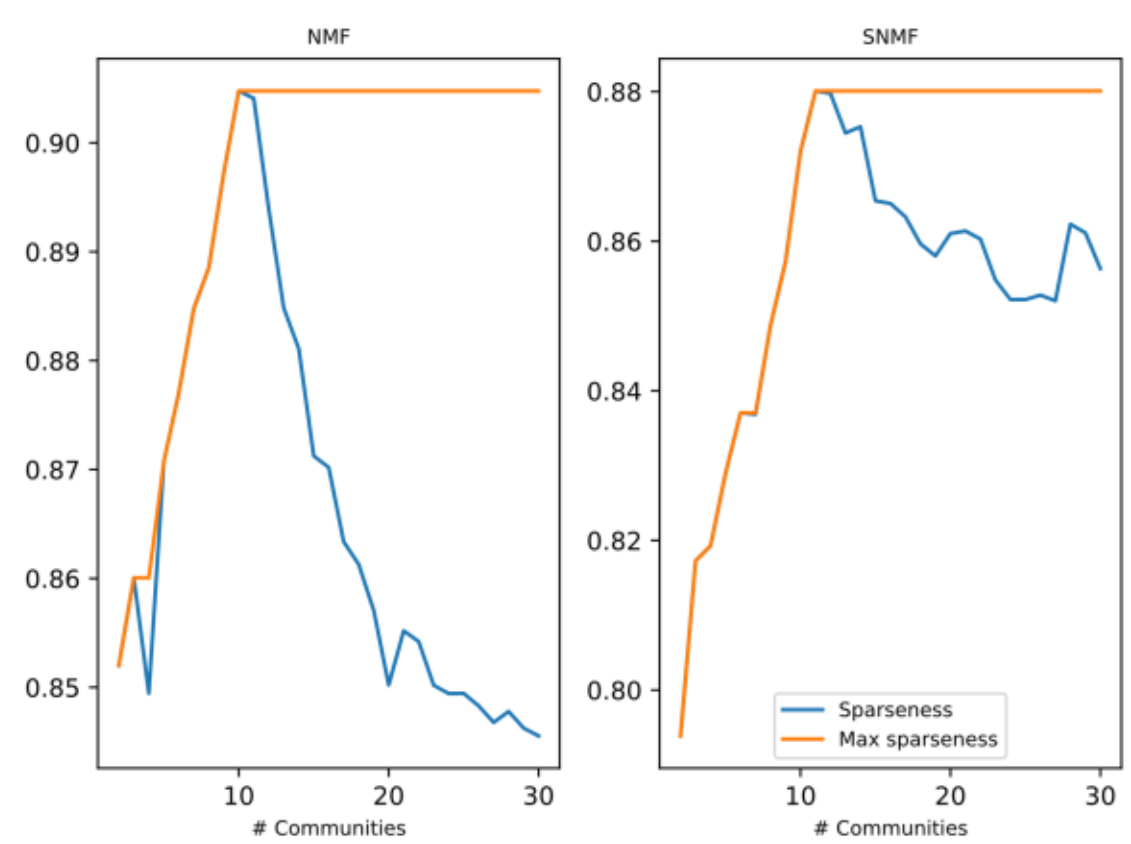

Fig. 3. Sparseness *vs.* number of communities using NMF in Eq. (7) and SNMF in Eq. (15) on the *American college football* network. The Sparseness curve indicates the average sparseness of *H* for various NMF/SNMF iterations while the Max Sparseness curve plots the maximum sparseness value from the initial iteration up to the current iteration. When the maximum sparseness gets close to 1, it becomes stable. At that point, NMF/SNMF iterations reach or get close to the representative number of components which we use as the number of communities. The SNMF sparseness is controlled while NMF sparseness decreases dramatically when NMF iterations surpass the number of communities in the network.

To control the degree of sparsity enforced, Hoyer proposed a measure for sparseness of a vector [27]:

$$sparseness(h)=\frac{\sqrt{n}\ -(\sum|h_i|)\,/\,\sqrt{\sum {h_i}^2}}{\sqrt{n}-1}, \tag{18}$$

where $n$ is the dimensionality of vector $h$. The sparseness of a matrix can be computed by averaging the sparsity of its vectors. In the context of community detection Eq. (18) can be interpreted as follows: if all community membership values of a node represented by a vector $h\in\mathbb{R}^k$ (for $k$ communities) are equal, the value of the $sparseness(h)$ becomes zero which means that the node is equally distributed across all communities. If only one value is non-zero, the $sparseness(h)$ becomes 1 (sparsest, hard clustering) and the node belongs to only one community; otherwise, the $sparseness(h)$ takes a fuzzy value indicating the degree of sparsity for a node belonging to several communities.

Iterating from $k' = 2$ to $k' = k_{max}$ where $k_{max}$ is the maximum number of possible communities; we run SNMF in Eq. (15) and compute the average sparseness over all columns of $H$ for each iteration. We keep track of the highest average sparseness using Eq. (18) and stop iterating when this average sparseness is nonincreasing for 10 iterations (to make sure that there is no other number of components that yield a higher sparseness). This parameter is determined by experiments but can be changed (increase for more confidence to check if there might be a higher value of sparseness or reduce to improve the speed). This indicates that we can use the number of components corresponding to the maximum sparseness of $H$ to effectively represent the entire network. For this reason, we use them as the estimated $k'$ as shown in Fig. 3 for the *American college football* network [55].

For the entire network of the *American college football*, 11 communities are correctly estimated when our approach is used with the SNMF in Eq. (15) and 10 communities when the same approach is used with the NMF in Eq. (7). This emphasises the superiority of SNMF over NMF. In fact, NMF solutions are not unique [56] [27]. Any matrix $X$ such that $WX \geq 0$ and $X^{-1}H \geq 0$ also provides a solution which causes unstable results that affect the community detection.

**Algorithm 3. Estimating the number of communities**

**INPUT**: Graph $G_s = (V_s, E_s)$
$A$ = adjacency matrix of $G_s$ in $n_s \times n_s$
**INITIALIZE:** $k_{max} = n_s/4$, $\beta = 1e-4$
    $x = 0.8$ #start with a good sparseness
    $counter = 1$ #counts of non-increasing sparseness
    $k' = 1$ #initial number of communities
    $finalH = \text{ones}(1, n_s)$ #start with a single community
**for** $temp_{k'}$ in range $(2, k_{max})$ **do**:
    $W, H = SNMF(A, temp_{k'})$ #using Eq. (15)
    $avg_s = 0; x_h = 0$ #for computing sparseness of $H$
    **for** $j$ in range $(0, n_s - 1)$**do:** #each node representation
        $x_h$ += sparseness$(h^{(j)})$ #using Eq. (18)
    **end for**
    $avg_s = {x_h}/{temp_{k'}}$ #compute the sparseness of $H$
    **if** $avg_s \leq x$ **do:** #sparseness not improving
        $counter$+= 1
    **else:** #the sparseness has improved
        $k' = temp_{k'}; finalH$ = Normalize$(H)$ #Eq. (13)
        $x = avg_s$; $counter = 1$ #reset the counter
    **end if**
    **if** $counter == 10$ **do:**#sparseness still not improving
        break
    **end if**
**end for**
**RETURN** $k', finalH$

To our knowledge, our method of estimating the number of communities has never been used in the literature and is superior to our previous MLC approach. We keep the sparseness parameter very small as high values increase the approximation error and the corresponding components may not represent the network. Algorithm 3 would be slow when it is used on a large network due to SNMF decomposition. This explains why the initial sampling in Algorithm 2 is required to speed up Algorithm 3. The complexity depends on the number of clusters in the sampled subgraph.

### C. Detecting Multiple Local Communities

**Algorithm 4. The overall S-MLC algorithm**

**INPUT**: Graph $G = (V, E)$, seed $s$
coms = [ ] # create an empty list of communities of $s$
$G_s$ = Algorithm2$(G, s)$
$k', H$ = Algorithm3$(G_s)$
$\theta = 1/|k'|$
$H\,[H \geq \theta] = 1$
**for** $i$ in range $(0, k' - 1)$ **do**
    $com = [v$ in nodes whose index in $H(i, :) = 1]$
    **append** $com$ to $coms$ if $s$ is in $com$
**end for**
**RETURN** $coms$

This is the final step of S-MLC for detecting multiple local communities for a specific seed. Given a community membership vector $h^{(i)} \in \mathbb{R}^{k'}$ from $H$ for a node $v_i$, we want to assign $v_i$ to its communities. Let $\theta = 1/|k'|$ be the average community membership probability for every node. A node that belongs to all $k'$ communities would be represented by a value that is equivalent to $1/|k'|$ in each of the $k'$ communities. We assign a node $v_i$ to community $j$ if $h_i^{(j)} \geq \theta$. This operation can be performed faster for all nodes using matrix operations, as shown in Algorithm 4. A node with a very high membership value (close to 1) in one community has less chance of being in any other community. A node whose community membership probabilities are equally distributed (equals to $\theta$) is assigned to all communities and overlaps appear easily.

Note that a user can force $\theta$ to be a specific value depending on the type of communities he (she) wants to detect. For example, if a user wants to detect communities where a node is allowed to be in no more than two communities, set $\theta = 0.5$. If one wants nodes to be in no more than 3 communities, then set $\theta = 0.33$; and if one wants communities whose members do not overlap with any other communities, set $\theta = 1$ or close to 1.

### D. Complexity

Approximate PageRank's time complexity depends on the graph size $n$ and the parameters used: $O\left(\frac{\log n}{\epsilon\alpha}\right)$ [15]. Extracting a biconnected component is affected by the maximum value between the number of nodes $n$ or the number of edges $m$ of the graph: $max\,(n, m)$ for both the time and space required [52]. Therefore, the complexity of local sampling is determined by the original graph size and the parameters used, $\epsilon$ and $\alpha$. Specifically, when $\alpha$ decreases, the more time it takes for sampling. For community number estimation, the complexity depends on the number of nodes $n_s$ in the sampled subgraph as the worst case of decomposing its adjacency matrix is $n_s/4$ times. This happens when the subgraph consists of very small clusters (≤ 4 nodes).

Normally a network may not consist of more than $n_s/4$ clusters as most clusters of real-world networks have sizes greater than 4 nodes. The community detection phase depends on the number of clusters obtained.

The overall complexity of our algorithm depends on $\epsilon$, $\alpha$ which determine the size of $G_s$ and the rest is $O(n_s)$.

## V. Experiments

All the experiments are done on a laptop with a processor: i5 @ 1.70 GHZ 1.70 GHZ, RAM: 8 GB and a 64-bit Windows operating system. All the algorithms are implemented in Python.

We first evaluate the local sampling, followed by community number estimation which we compare with the modularity maximization, then we compare the proposed local community detection approaches with MLC [19] and M-LOSP [21]. As there are few algorithms for multiple local community detection, we also run DEMON [13], one of the popular global overlapping community detection algorithms, on the sampled subgraph to find approximate local communities, denoted as L-DEMON. For networks with ground-truth communities, we compare the detected communities with the ground-truth communities using $F_1$ and $F_2$ scores. A high $F_1$ score is important because it focuses on both good precision and good recall. This indicates the algorithm's ability to discard irrelevant nodes such as innocent people in criminal detection and find all relevant nodes (criminal people). $F_2$ score is high when the recall is relatively high as less attention is given to the precision. In criminal detection, if all criminals are found, $F_2$ would be high even if some innocents are included as criminals.

NMI [57] [8] is another measure often used to evaluate the community detection quality but it is suitable for evaluating global communities and not local communities. This is because any two community assignments (ground-truth and detected) should have the same number of nodes $n$ in order to be compared using NMI (requires $n$ ). In local community detection, there is a detected portion and the rest of the network, and some nodes in the ground-truth communities may not appear in the detected communities and vice-versa. For networks without ground-truth communities, we compute the conductance to evaluate the quality of the detected communities.

### A. Datasets

#### 1) Artificial Networks

We use the LFR benchmark networks [58] which simulates the characteristics of real-world networks as opposed to the GN benchmark [59], which assumes that all nodes have the same degree and all communities are of the same size. Eight unweighted, undirected LFR networks are generated by modifying the mixing parameter $\mu$, minimum community size $C_{min}$, maximum community size $C_{max}$, number of overlapping nodes *on* and number of memberships of the overlapping nodes *om*. Each network consists of 1000 nodes ($n_s$) with different number of edges ($m_s$). We choose the average degree $d$ = 5 and the highest degree $d_{max}$ = 15.

TABLE I
Real-world Benchmarks

| Network | *n* | *m* | *k* | average degree |
|---|---|---|---|---|
| Zachary karate club | 34 | 78 | 2 | 4.59 |
| Dolphin | 62 | 159 | 4 | 5.13 |
| American college football | 115 | 613 | 10 | 10.66 |
| Amazon | 334,863 | 925,872 | 75,149 | 5.53 |
| DBLP | 317,080 | 1,049,866 | 13,477 | 6.62 |
| ego-Facebook | 4,039 | 88,234 | - | 43.69 |
| gemsec Politician | 5908 | 41,729 | - | 14.12 |
| Gemsec TvShow | 3,892 | 17,262 | - | 8.87 |
| musae-Facebook | 22,470 | 171,002 | - | 15.22 |

*n*: number of vertices; m: number of edges;
*k*: number of ground-truth communities;
-: no ground-truth communities.

#### 2) Real-world Networks

We use a combination of small and large networks often used to evaluate community detection algorithms. The statistics of the networks are shown in Table I.

*Zachary karate club* is a social network of friendships among 34 members of a karate club in a US university in the 1970s [45]. A conflict between the club administrator and the instructor over the price of a karate lesson led the network to be split into two groups.

*Dolphin* is a network of 62 dolphins observed between 1994 and 2001 in Doubtful Sound, New Zealand. An edge connects two dolphins that were seen together 'more often than the expected chance'. The network consists of two main communities although one community can be split into three sub-communities [60].

*American college football* is a network of games between Division I colleges in 2000 [59]. Colleges are grouped into conferences and each conference is a ground-truth community. Colleges (teams) are nodes and edges denote games between the teams.

*Amazon* is a network with ground-truth communities consisting of products which are often purchased together [1]. *DBLP* is a co-authorship network with ground-truth communities consisting of authors who published on the same journal or conference [1]. Both *Amazon* and *DBLP* have a minimum ground-truth community size of three nodes and all their ground-truth communities are defined based on meta data.

*Ego-Facebook* is a network of friendships on Facebook collected from 4039 users with a mobile app [1]. *Gemsec-Politician, Gemsec-TvShow* and *Musae-Facebook* are networks representing verified Facebook pages of different categories. Nodes represent pages while edges indicate mutual likes between pages [1]. These networks do not have ground-truth communities.

### B. Local Sampling Results

We evaluate the quality of local sampling on large real-world networks only as the community detection on small networks

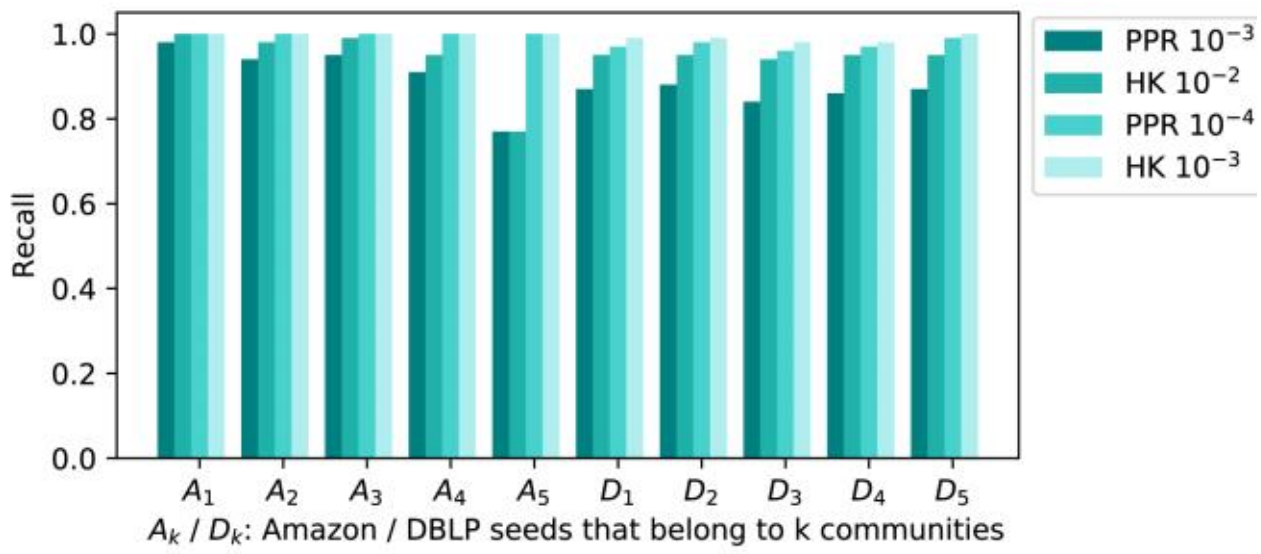

Fig. 4. PPR and HK sampling recall on *Amazon* and *DBLP*. HK returns all nodes in the ground-truth communities of the seed for both values of $\varepsilon$ used. For PPR, the recall is slightly lower compared to HK. For both algorithms, as we decrease $\varepsilon$ we get a higher recall.

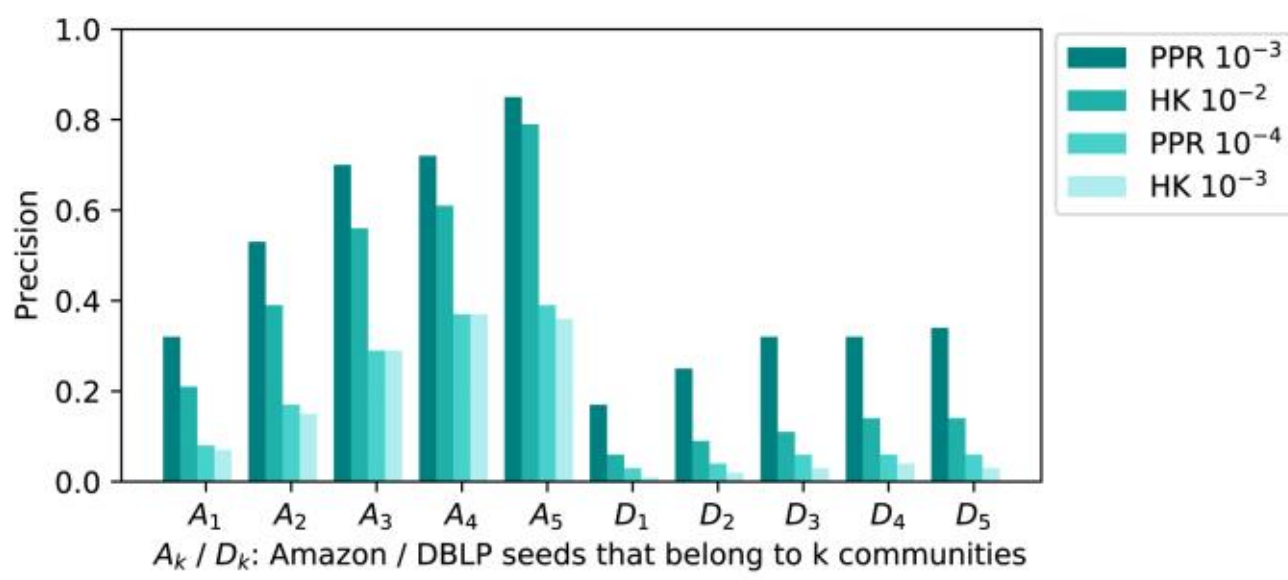

Fig. 5. PPR and HK sampling precision on *Amazon* and *DBLP*. PPR has a higher precision than HK on both datasets for all 1000 seeds used in total. PPR returns less irrelevant nodes during sampling compared to HK. On *Amazon*, seeds that belong to 5 communities have a high precision as most of their communities consist of a high overlap and sometimes, they may look like a single community.

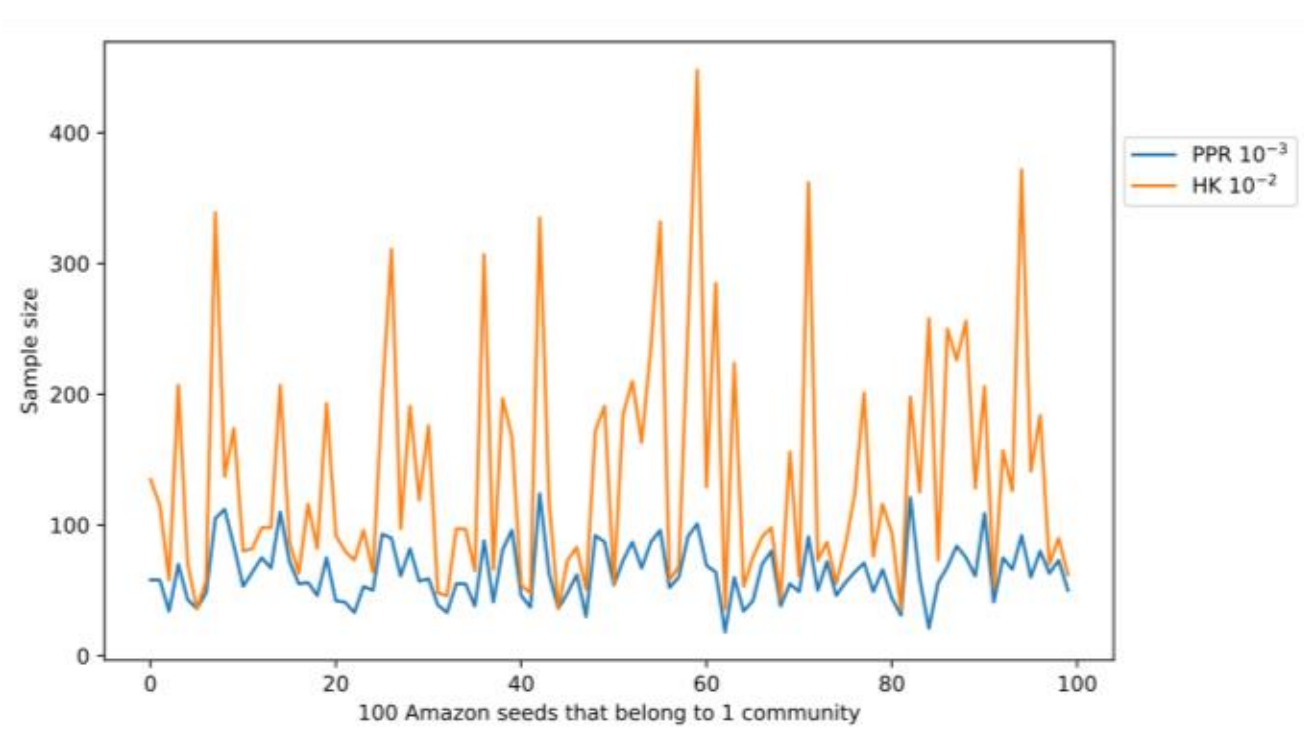

Fig. 6. An overview of sample sizes of HK $10^{-2}$ and PPR $10^{-3}$.

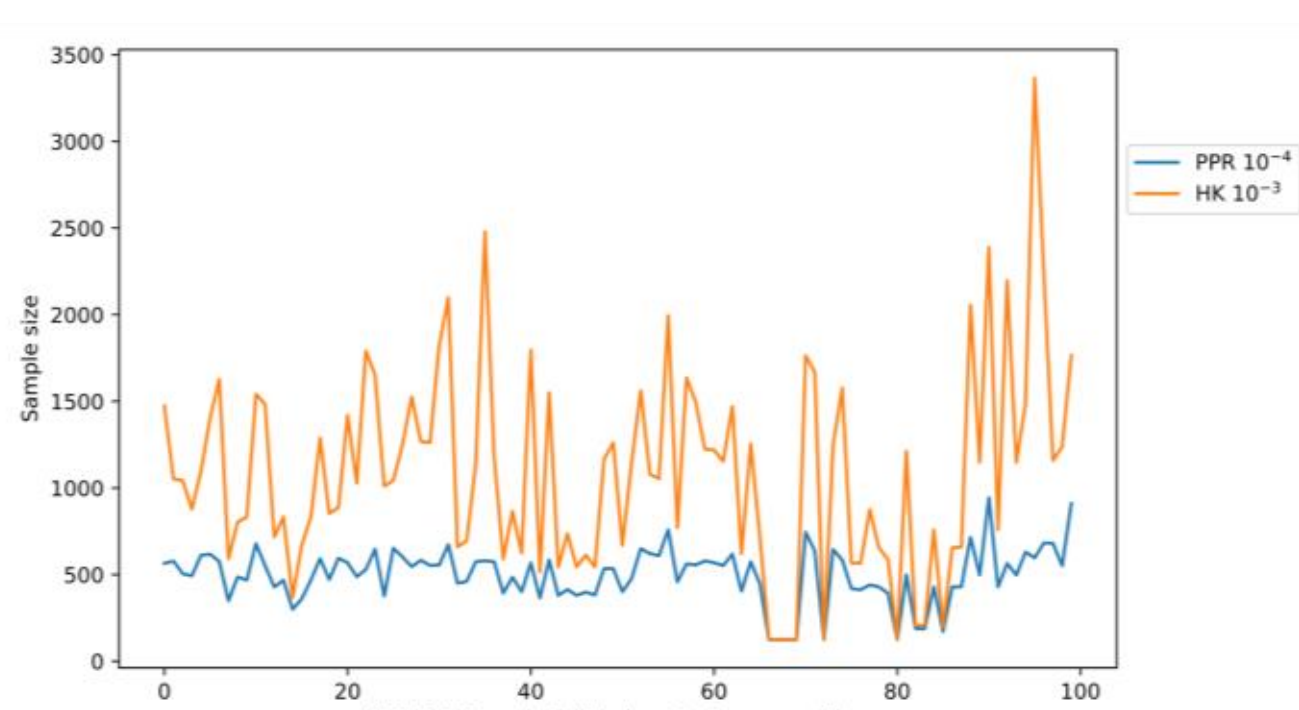

Fig. 7. An overview of sample sizes of HK $10^{-3}$ and PPR $10^{-4}$.

TABLE II
NUMBER OF COMMUNITIES OF SMALL REAL-WORLD NETWORKS

| Network | $k$ | $k'$ | $k_{mod}$ |
|---|---|---|---|
| Zachary karate Club | 2 | **2** | **2** |
| Dolphins | 2 | **2** | 3 |
| American college football | 11 | **11** | 10 |

$k$: number of ground-truth communities;
$k'$: number of communities estimated by the sparseness approach;
$k_{mod}$: number of communities detected by modularity maximization.

TABLE III
NUMBER OF COMMUNITIES ON ARTIFICIAL NETWORKS

| Network | LFR parameters | | | | | $n$ | $m$ | $k$ | $k'$ | $k_{mod}$ |
|---|---|---|---|---|---|---|---|---|---|---|
| | $\mu$ | $C_{min}$ | $C_{max}$ | *on* | *om* | | | | | |
| $G_1$ | 0 | 10 | 30 | 0 | 0 | 1000 | 2778 | 57 | 61 | 53 |
| $G_2$ | 0 | 100 | 200 | 0 | 0 | 1000 | 2813 | 6 | **6** | 5 |
| $G_3$ | 0 | 3 | 20 | 20 | 2 | 1000 | 2667 | 112 | 115 | **112** |
| $G_4$ | 0 | 20 | 200 | 20 | 2 | 1000 | 2769 | 9 | **9** | 8 |
| $G_5$ | 0.1 | 20 | 200 | 20 | 2 | 1000 | 2745 | 13 | **13** | 12 |
| $G_6$ | 0.1 | 20 | 200 | 20 | 3 | 1000 | 2721 | 10 | **10** | **10** |
| $G_7$ | 0.1 | 20 | 200 | 20 | 4 | 1000 | 2756 | 11 | **11** | 10 |
| $G_8$ | 0.1 | 20 | 200 | 20 | 5 | 1000 | 2837 | 18 | 17 | 16 |

$\mu$: mixing parameter; $C_{min}$: minimum community size;
$C_{max}$:maximum community size; on: number of overlapping vertices;
*om*: number of memberships of the overlapping vertices.

can be done without local sampling. Fig. 4 and Fig. 5 show the results obtained on *Amazon* and *DBLP* datasets using PPR approximation with $\alpha = 0.99$ and $\varepsilon \in \{10^{-3}, 10^{-4}\}$ and HK approximation with $\varepsilon \in \{10^{-3}, 10^{-4}\}$ and $t \in \{80, 40\}$ respectively. Each category ($A_k$ *or* $D_k$) indicates the average results computed over 100 seeds.

Overall, the precision is higher on *Amazon* than *DBLP* and PPR has a higher precision than HK while HK has a higher recall than PPR. On both PPR and HK, the lower the $\varepsilon$, the lower the resultant precision and the higher the resultant recall. The sampling results indicate that if we use HK sampling, most nodes we need will be sampled along with many irrelevant nodes that will need to be discarded during the community detection phase. If we use PPR, we may not sample all nodes we need but we get fewer irrelevant nodes that need to be discarded during the community detection phase. As the precision gets lower, the community detection becomes much harder as a result of having many nodes to discard. On *DBLP*, we have a very high recall compared to the precision which predicts a harder community detection.

The sampling results differ based on the characteristics of the networks. For example, on *Amazon*, books can belong to literature and fiction as one community. From these books, there are books for teens and young adults that can form a second community while some books may be related to history (historical fiction) and form the third community. In this case, many books overlap in various communities and this is the same for other products. As a result, the overlapping portion is denser compared to *DBLP* and when you take an *Amazon* seed, most of its community members are easily sampled with an improved precision compared to *DBLP*. This explains the reason why the sampling precision is higher for seeds belonging to many communities than for seeds belonging to a single community, as shown in Fig. 5.

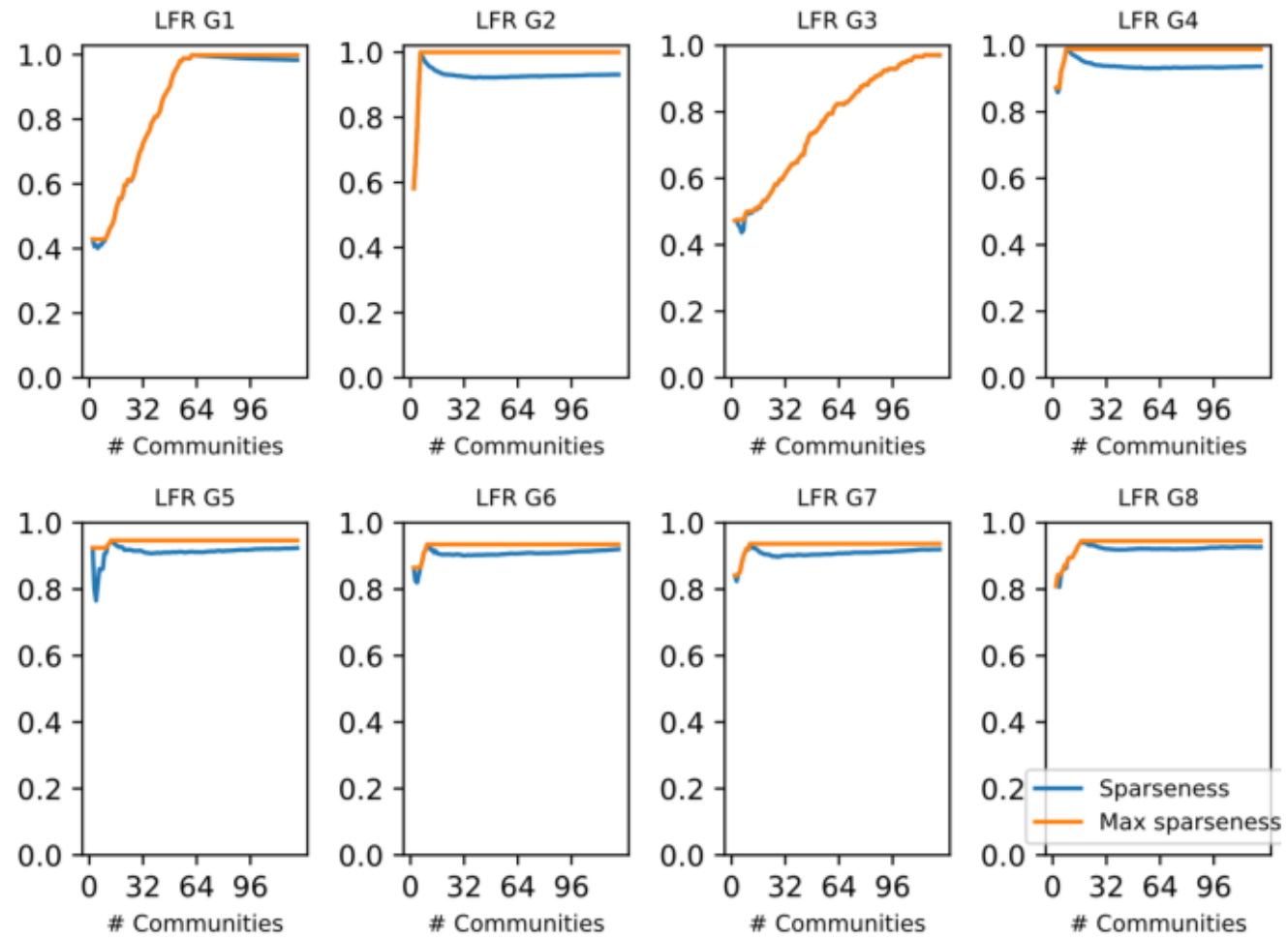

Fig. 8. Sparseness versus number of communities in LFR networks. The non-increasing sparseness correlates with the ground-truth number of communities for most networks. The sparseness becomes stable when greater than 0.8.

Fig. 6 and Fig. 7 show a comparison of PPR and HK. HK $10^{-2}$ samples more nodes compared to PPR $10^{-3}$, and HK $10^{-3}$ sample a larger number of nodes compared to both PPR $10^{-3}$ and PPR $10^{-4}$.

### C. Results on Estimating the Number of Communities

We use small real-world networks (Table II) whose number of communities is known and eight LFR networks summarized in Table III.

Eight LFR networks are used to estimate their number of communities. As these networks are small, we set $G_s = G$ so that we can estimate the number of communities in the entire network. Fig. 8 shows the correlation between the sparseness and the number of communities. The non-increasing sparseness correlates with the ground-truth number of communities on graphs that consist of a few communities ($G_2$, $G_4$, $G_5$, $G_6$, $G_7$) compared to graphs with many communities ($G_1$ and $G_3$). The sparseness-based approach accurately estimates the number of communities in $G_2$, $G_4$, $G_5$, $G_6$, $G_7$. The modularity approach accurately estimates all the 112 communities in $G_3$ and $G_6$. In practice, a seed belongs to less than 5 communities and our algorithm is more accurate than the modularity-based approach when estimating the number of communities in such cases.

### D. Local Community Detection Results

We compare S-MLC with M-LOSP, MLC and L-DEMON using $F_1$ and $F_2$ scores. S-MLC uses PPR with $\varepsilon = 10^{-3}$ for local sampling due to the improved precision as compared to HK (Fig. 5). Fig. 9 and Fig. 10 show the results on artificial networks. Fig. 11 and Fig. 12 show the results on small real-world networks while Fig. 13 and Fig. 14 show the results on large real-world networks. Fig. 15 provides insights of the quality of the communities detected by S-MLC based on the conductance measure (Eq. 1). Fig. 16 compares the conductance results of different algorithms on datasets without ground-truth communities.

For the artificial networks summarized in Table III, we select 100 seeds that belong to one community for each of $\{G_1, G_2\}$ Then for each of $\{G_3, G_4, G_5\}$, we select all seeds that belong to two communities, and for $\{G_6, G_7, G_8\}$, we select all seeds that belong to three, four and five communities respectively.

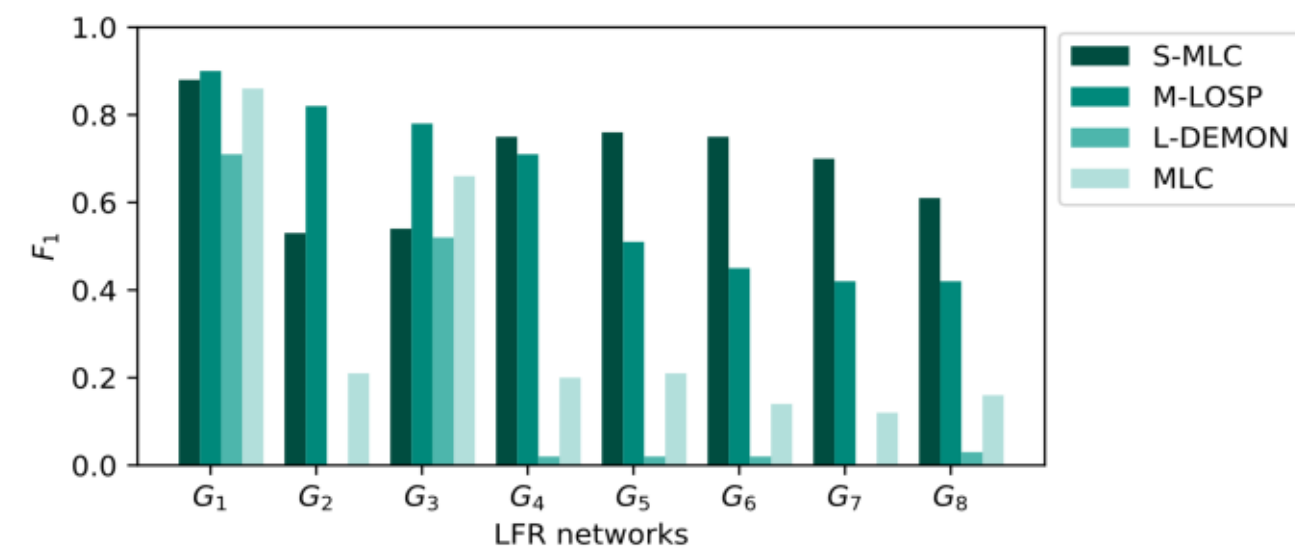

Fig. 9. $F_1$ results on LFR networks. S-MLC outperforms MLC, M-LOSP and L-DEMON on most networks except on $G_2$ and $G_3$. $G_3$ consists of very small communities while $G_2$ consists of large communities with a good community structure. DEMON's approach does not guarantee that each node will be assigned to a community. Thus, most detected communities do not include the seed and the averages of $F_1$ scores are very low.

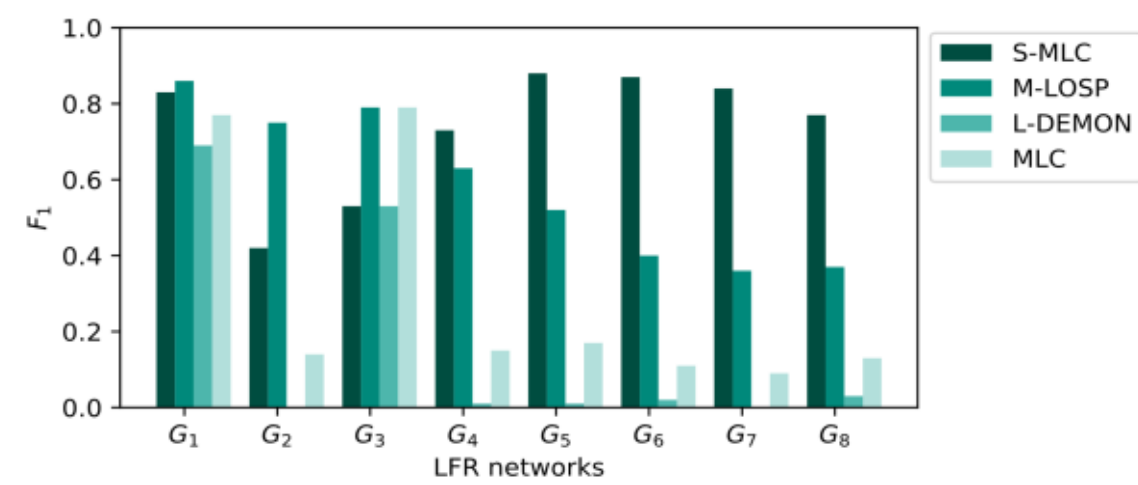

Fig. 10. $F_2$ results on LFR networks. S-MLC outperforms the other three algorithms of most networks except on $G_2$ and $G_3$.

Fig. 9 and Fig. 10 show the $F_1$ and $F_2$ scores obtained respectively. All the algorithms easily detect small single communities in $G_1$. M-LOSP and S-MLC outperform other algorithms in detecting large single communities in $G_2$. M-LOSP and MLC outperform the rest of the algorithms in finding small overlapping communities in $G_3$ while S-MLC dominates other algorithms in detecting large overlapping communities in $\{G_4, G_5, G_6, G_7, G_8\}$.

For some seeds, L-DEMON gets very low scores. For such seeds most returned communities do not include the seed and in such cases, the corresponding $F_1$ and $F_2$ scores become zero. For each node $v$ in the sampled subgraph, L-DEMON finds its ego network (direct neighbors of $v$), then removes $v$ from the ego network and applies Label Propagation algorithm [13] on the remaining nodes to find communities. Label Propagation assigns a node $v$ to a community $C$ if the maximum number of neighbors of $v$ belong to $C$. When $v$ is not densely connected, the Label Propagation operates on a very small subgraph and returns very small communities often consisting of two or three nodes each. Such communities are discarded by L-DEMON when they are no greater than a particular threshold (default threshold is 3). Larger communities are merged with or appended to previously detected communities. To sum-up, although L-DEMON finds communities in an entire subgraph given as the input, there is no guarantee that all nodes will appear in the detected communities.

On small real-world networks, S-MLC also outperforms the rest of the algorithms (Fig. 11 and Fig. 12). As we did not use a sampling on such networks (for all algorithms), the results indicate the superiority of the community estimation approach and community detection approach of S-MLC over the ones used in MLC as indicated by $F_1$ and $F_2$ scores.

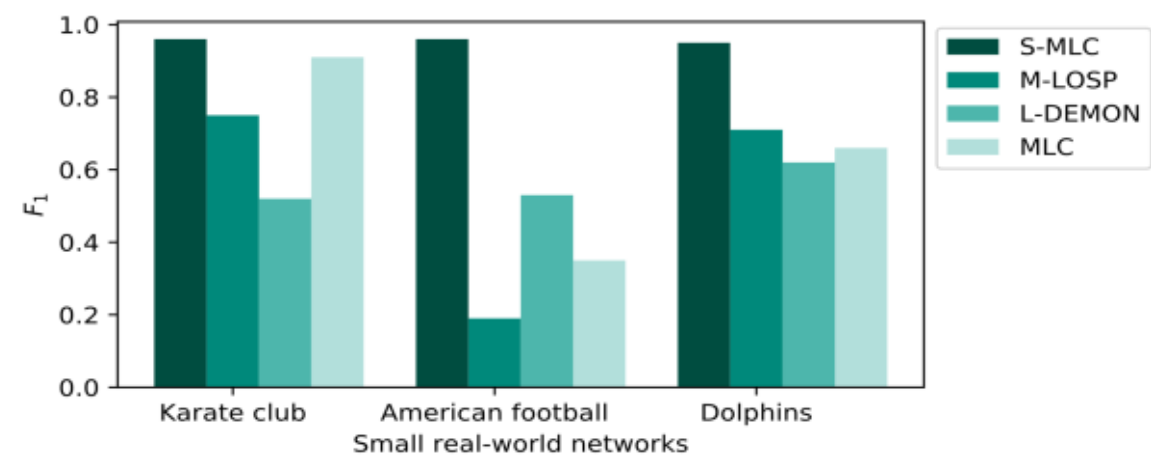

Fig. 11. $F_1$ results on small real-world networks. S-MLC outperforms the other three baselines.

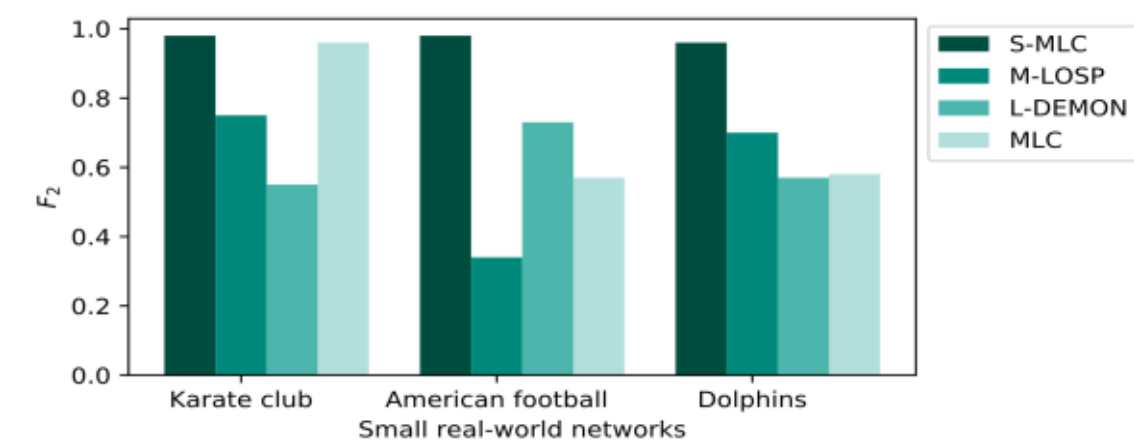

Fig. 12. $F_2$ results on small real-world networks. S-MLC outperforms the other three baselines.

On large real-world networks, S-MLC performs well on *Amazon* and M-LOSP performs well on *DBLP* as shown in Fig. 13 and Fig. 14. Overall, S-MLC performs well when a good sample is available. For instance, a very good performance on seeds for five communities ($A_5$) is based on a high sampling precision. S-MLC uses PPR-$10^{-3}$ for local sampling. This is the case of small networks and *Amazon*. On *Amazon*, in general, S-MLC outperforms M-LOSP, followed by L-DEMON and MLC. On *DBLP*, in general, M-LOSP is the best, S-MLC and L-DEMON are competitive, followed by MLC. We see that when there is a sample with many irrelevant nodes that need to be discarded, such as *DBLP* (Fig. 5), S-MLC is still able to discard most of the irrelevant nodes and uncover more than 50% of the relevant nodes as shown by $F_1$ and $F_2$ scores.

In summary, S-MLC shows favorable accuracy as compared with the other three baselines. We further see that all communities detected by S-MLC have a good conductance less than 0.5 on average (Fig. 15). A small conductance indicates a good community structure and the worst community structure has a conductance of 1. On *Amazon*, the quality of the detected communities is lower than the quality of the ground-truth communities based on their conductance. Most of the time, this is caused by detected communities larger than the ground-truth communities where additional nodes reduce the quality. On *DBLP*, it is opposite. The ground-truth communities are very large with a low community quality (high conductance) while the detected communities are smaller and more compact as indicated by lower conductance compared to the ground-truth communities. This kind of interpretation is specifically useful when we do not know the ground-truth communities as illustrated in Fig. 16. This means that users in the *ego-Facebook* network tend to form compact groups of friends where a user

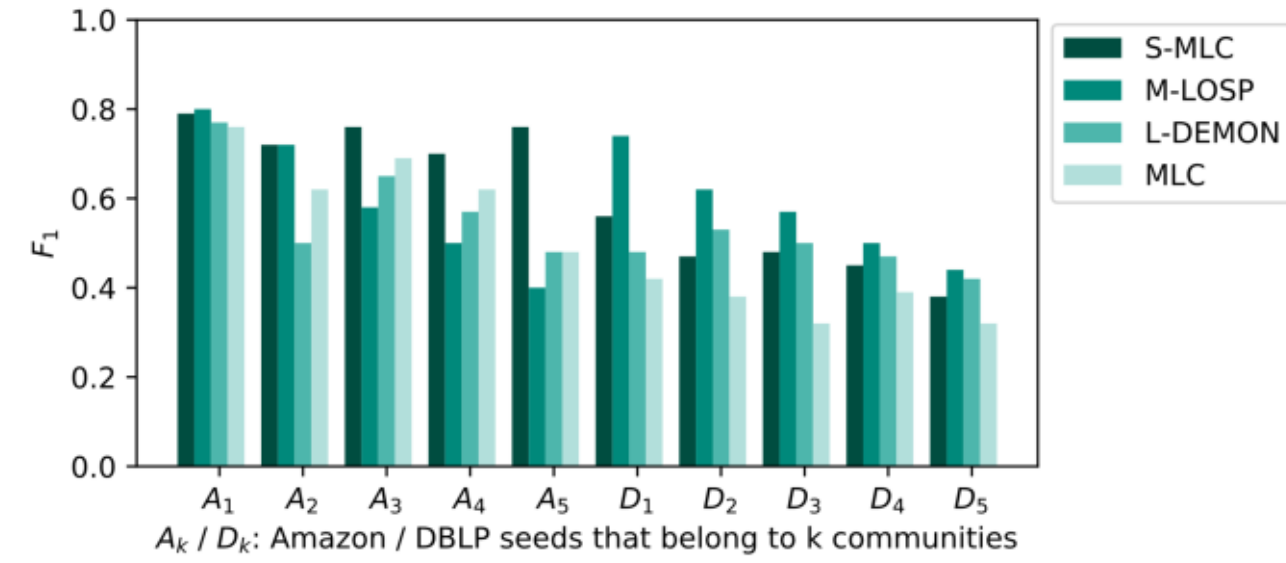

Fig. 13. $F_1$ results on large networks. In general, S-MLC outperforms M-LOSP L-DEMON and MLC on *Amazon*, especially for $A_5$ which contains more local communities. On *DBLP*, S-MLC outperforms MLC with a performance close to L-DEMON's. On DBLP, M-LOSP is the best, followed by S-MLC, L-DEMON and MLC.

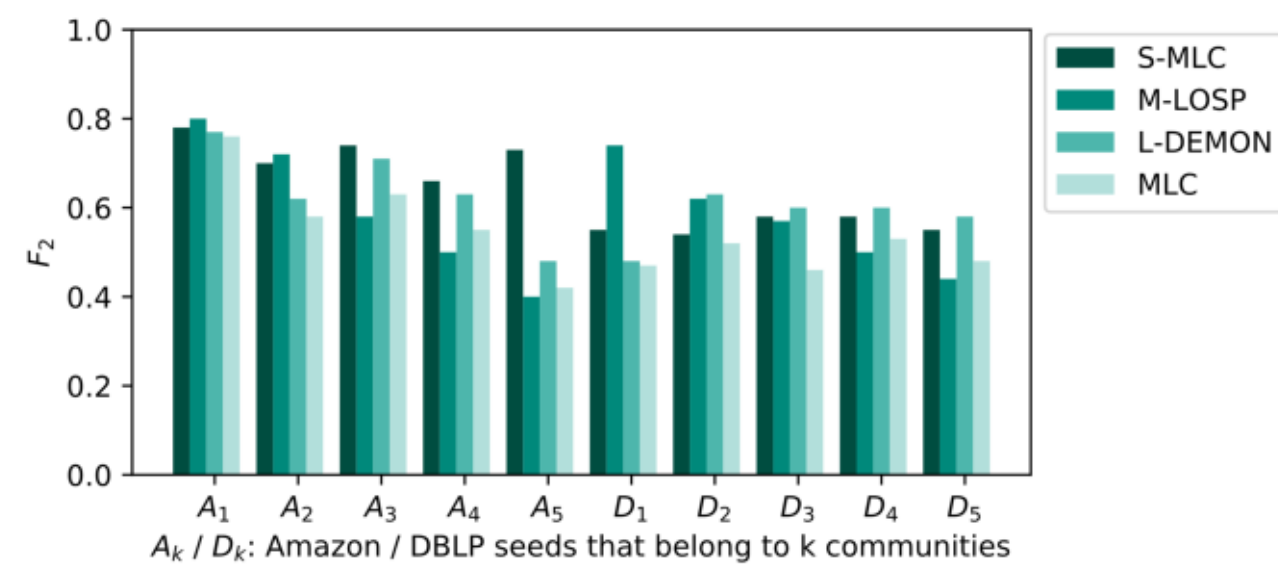

Fig. 14. $F_2$ results on large networks. Overall, $F_2$ are higher than $F_1$ results. In general, S-MLC outperforms M-LOSP, L-DEMON and MLC on *Amazon*. On DBLP, L-DEMON has a higher $F_2$ score compared to the other three methods.

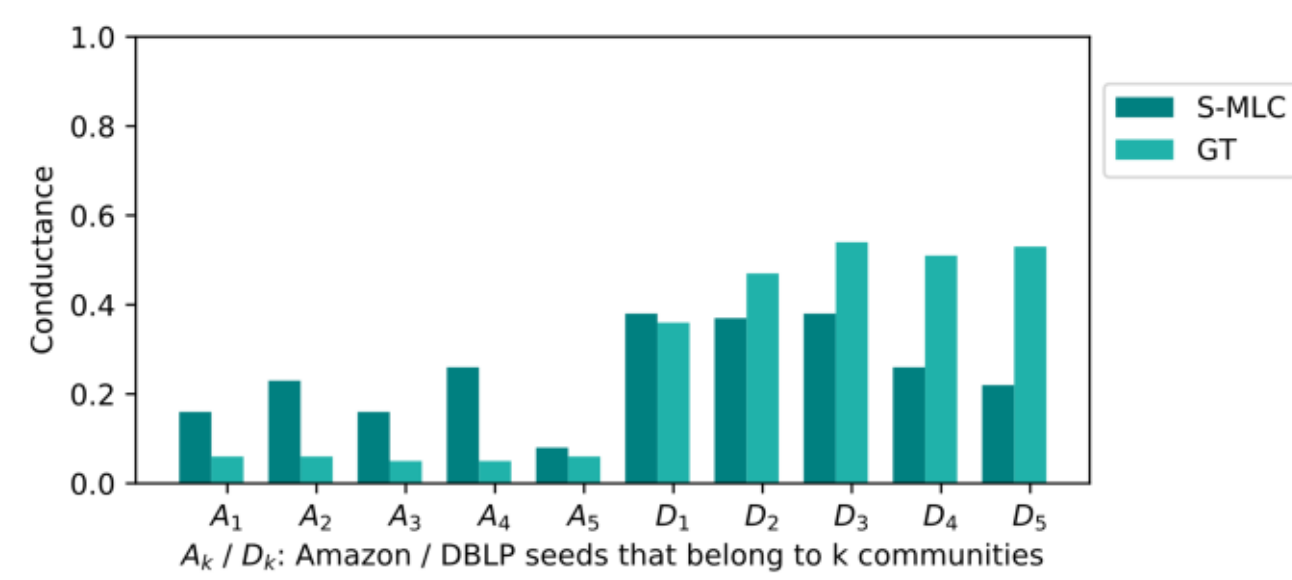

Fig. 15. Conductance of the detected communities compared to the conductance of the ground-truth communities. GT denotes the ground-truth communities.

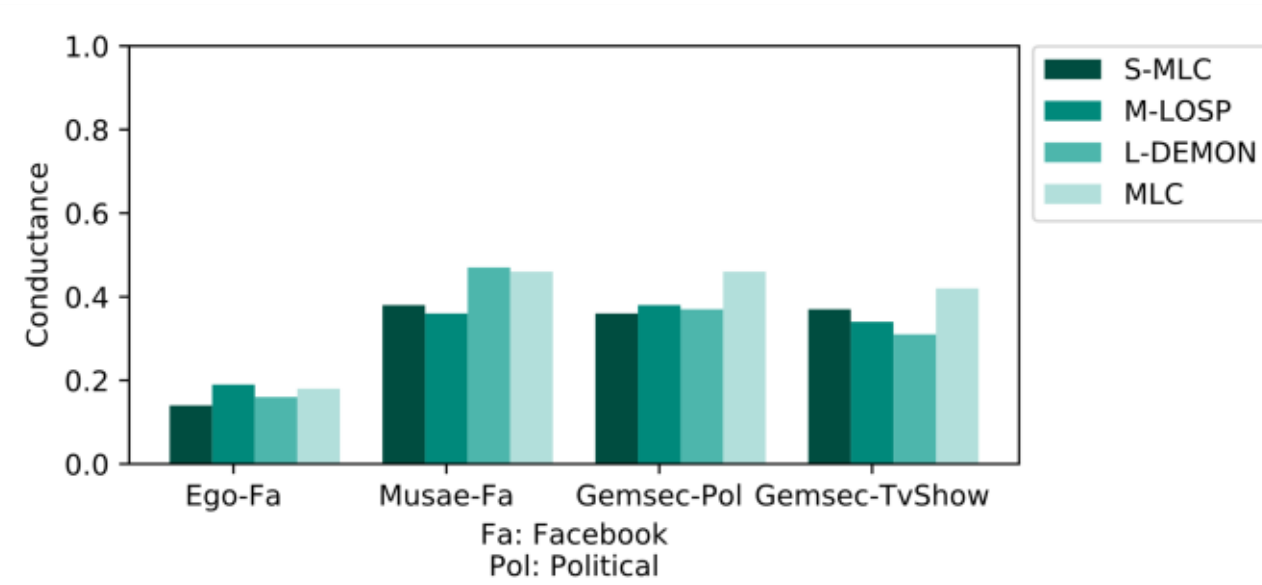

Fig. 16. The conductance of the detected communities for different algorithms on various Facebook datasets. The conductance is low on the *ego-Facebook* network (Ego-Fa) which represents a good community structure while the conductance is high on datasets that represent mutual likes between pages.

maybe a friend of almost all members in a same community.

For Facebook pages, one page may like many pages which do not like each other and their community structure is not well structured as the one formed by Facebook users.

## VI. CONCLUSION

As opposed to existing local community detection methods that focus on detecting a single community supervised by some exemplary seeds, we assume that a seed may belong to multiple communities and propose S-MLC, a new method of detecting multiple local communities. The method is based on three steps which can be used independently by other community detection algorithms. For instance, our approach of estimating the number of communities in a network can be used in other algorithms that require the number of communities as prerequisites.

We evaluated S-MLC using real-world and artificial networks and experiments showed favorable accuracy on S-MLC, which in general outperforms the three state-of-the-art baselines on artificial networks, small real networks and large real network of *Amazon*, only having exceptions on large real network of *DBLP*. The evaluation also showed that the community detection results depend highly on the sampling quality. A good sampling generates all nodes of the target communities with no or few irrelevant nodes to be discarded during community detection. The sampling also has an impact on the complexity of our algorithm as subsequent community detection tasks depend on the size (number of nodes) of the subgraph $G_s$. PPR was selected over HK for sampling as PPR generates samples with a higher precision compared to HK.

For networks without ground-truth communities, the smaller the conductance of the detected communities, the better are the obtained communities and S-MLC generated communities with an average conductance less than 0.5.

## ACKNOWLEDGMENT

We acknowledge the Ministry of Commerce (MOFCOM) of the People's Republic of China and the University of Rwanda for sponsoring the studies of the first author.